# Comment on "Consequences of the single-pair measurement of the Bell parameter"

Marian Kupczynski [*]
*Département de l'Informatique et Technologie, UQO, Case Postale 1250, Succursale Hull, Gatineau, Quebec, Canada J8X3X7*

Genovese and Piacentini [Phys. Rev. A **111**, 022204 (2025)] claim that, in a recent experiment [S. Virzì *et al.*, Quantum Sci. Technol. **9**, 045027 (2024)], the Bell parameter was measured on a single pair of photons, thus challenging several conclusions and discussions on the meaning of Bell inequalities as well as certain quantum mechanical interpretations. We explain that the parameter measured in the experiment of Virzì *et al.* is not the Bell parameter *S*, which has been discussed and estimated in many loophole-free tests. Therefore, this experiment neither challenges our understanding of Bell tests nor allows having doubts about Bohr complementarity and contextuality.

In their article, Genovese and Piacentini [1] present a comprehensive discussion of recent literature on Bell tests and discuss implications of the experiment by Virzì *et al.* [2].

They start with the definition of the Bell parameter. "If one considers the correlations between Alice's and Bob's measurements performed randomly choosing between two settings and respectively dubbed $A_i$ and $B_j$ $(i, j = 1, 2)$, by averaging on multiple experiments one can write the correlation parameter

$$\mathcal{B} = \langle A_1B_1 + A_1B_2 + A_2B_1 - A_2B_2\rangle. \tag{1}$$

Then, it can be demonstrated that $|\mathcal{B}| \leqslant 2$ for every LHVT the so-called Clauser-Horne-Shimony-Holt (CHSH) inequality but can reach $2\sqrt{2}$ in quantum mechanics."

This definition requires an explanation. In (1), $B = \langle C\rangle$, where $C$ is a random variable being a function of four jointly distributed binary random variables $A_1$, $A_2$, $B_1$, and $B_2$ [3–7]. The values of $C$ for each trial can only be evaluated using an $N\times 4$ spreadsheet in which outcomes $(\pm1, \pm1, \pm1, \pm1)$ of four measured random variables are displayed [8]. Then, for all the finite samples the estimates of $|\mathcal{B}|$ are smaller than or equal to 2 and the inequality $|\mathcal{B}| \leqslant 2$ cannot be violated.

Neither in an ideal Einstein-Podolsky-Rosen-Bohm experiment [9] nor in Bell tests [10] do joint probability distributions of $A_1$, $A_2$, $B_1$, and $B_2$ exist and Bell knew it very well. Therefore, the Bell parameter evaluated in quantum mechanics and estimated in Bell tests is not $\mathcal{B}$, as it was defined in (1), but $S$ [9]:

$$S = \langle A_1B_1\rangle + \langle A_1B_2\rangle + \langle A_2B_1\rangle - \langle A_2B_2\rangle. \tag{2}$$

The CHSH inequality is not $|\mathcal{B}| \leqslant 2$ but $|S| \leqslant 2$. Bell did not notice [3–8] that in the proofs of Bell and CHSH inequalities the existence of a joint probability distribution of $A_1$, $A_2$, $B_1$, and $B_2$ was tacitly assumed.

In Bell tests pairwise expectation values in (2) can only be estimated in four different random experiments; thus these estimates may violate $|S| \leqslant 2$. Four $N\times 2$ data spreadsheets cannot be reshuffled to form quadruplets [11] and the CHSH inequality is violated. Even if one assumes, as in the local realistic hidden-variable model [7–9], that experimental outcomes are predetermined, estimates of $S$ violate the CHSH inequality 50% of time [12,13]. Only if sample size $N$ increases does the violation of (2) become less significant.

In quantum mechanics every experiment is described by its dedicated context-dependent probabilistic model; thus

$$\langle A_iB_j\rangle = \mathrm{Tr}(\rho\hat{A}_i \otimes \hat{B}_j) \tag{3}$$

and $|S| \leqslant 2\sqrt{2}$. If in (1) one substitutes the random variables by operators $A_iB_j \to \hat{A}_i \otimes \hat{B}_j$, one obtains an operator $\hat{C}$ which is Hermitian but its physical meaning and how it could be directly measured is not clear because it is a sum of noncommuting operators. If one had access to a hypothetical device which implements the observable C, measurement outcomes would be nonseparable, i.e., it would be not possible for Alice and Bob to obtain a unique measurement outcome for each observable (for which we usually obtain the standard $\pm1$ value); instead a joint outcome would be obtained. Therefore, the individual pairwise expectations $\langle A_1B_1\rangle$, $\langle A_1B_2\rangle$, $\langle A_2B_1\rangle$, and $\langle A_2B_2\rangle$ could not be resolved from measuring $\mathcal{B}$, which remains consistent with the impossibility of resolving incompatible observables in a single shot. One may draw an analogy to resolving only the parity of two bits, which allows one to describe the joint outcomes, but it does not allow knowing the individual bitwise values. Similarly, knowing only the total energy of a quantum system, one does not have the information about its kinetic and potential energies. This is why there is no clear connection between measuring $\mathcal{B}$ and testing the CHSH inequality.

Nevertheless, if (3) is used to evaluate $|S|$, then according to quantum mechanics $|\langle\hat{C}\rangle| = |S|$ and this equality is used, as a convenient mathematical tool, to prove that $2\sqrt{2}$ (the

[*]Contact author: marian.kupczynski@uqo.ca

Tsirelson bound) is the maximal value of $S$ for any quantum state $\rho$ and for a large class of quantum operators [8,14,15].

According to contextuality-by-default approach, random variables should be labeled by an experimental context and a measured content [16]. Therefore, experimental data in Bell tests are described by four pairs of random variables $(A_{11}B_{11})$, $(A_{12}B_{12})$, $(A_{21}B_{21})$, and $(A_{22}B_{22})$ and *a priori* $|S| \leqslant 4$ [17]. Therefore, in Bell tests [10] we can only assess a plausibility of different probabilistic models or couplings [18]. In particular, we compare the quantum coupling

$$\langle A_{ij}B_{ij}\rangle = \mathrm{Tr}(\rho \hat{A}_i \otimes \hat{B}_j) \tag{4}$$

with the Bell-local realistic coupling

$$\langle A_{ij}B_{ij}\rangle = \int_{\Lambda} A_i(\lambda)B_j(\lambda)\rho(\lambda)d\lambda. \tag{5}$$

The meaning and implications of Bell tests are now well understood. Bell tests allowed rejecting, with high confidence, Bell-local and Bell-causal noncontextual hidden-variable models [10–13,17–19] proposed as probabilistic couplings like the coupling (5). They also confirmed the validity of Bohr complementarity and contextuality. The measuring instruments do not passively read preexisting values of physical observables but play an important and active role.

Genovese and Piacentini do not make a distinction between $\mathcal{B}$ and $S$ and claim that in the experiment [2], the Bell parameter was measured on a single pair of photons. Therefore, it challenges not only the explanation of Bell tests, reviewed above, but also several interpretations of quantum mechanics [20–22].

As we explained above, the Bell parameter is $S$ and not $\mathcal{B}$; thus it cannot be measured for a single pair of entangled photons. The $S$ contains four pairwise expectation values (2), each bounded by 1, which can be estimated only in dedicated experiments performed in four different incompatible experimental contexts.

The values of $\mathcal{B}$ and $S$ by the definition cannot exceed 4. Therefore, neither $\mathcal{B}$ nor S is measured in [2] but a different parameter, let us call it $S_{\mathrm{weak}}$, which can take values as large as $\pm 300$ (see Fig. 2 in [2]).

The experiment in [2] follows a completely different protocol than Bell tests. Polarization-entangled photons are generated in a Sagnac interferometer, fiber coupled, and then collimated in a Gaussian spatial mode. Then each of them undergoes two weak measurements in sequence, each realized by a pair of birefringent crystals preceded by a half waveplate, before impinging on a $24 \times 24$ single-photon avalanche detector array detecting the photon pairs with two-dimensional spatial resolution. Entangled photonic signals slightly deviate and Alice and Bob can extract the information on their polarization measurements, as well as on their cross correlations, from the coordinate set of $X_A$, $Y_A$, $X_B$, and $Y_B$ generated by the detection of each entangled photon [2].

It is true that for each quadruple $X_A$, $Y_A$, $X_B$, and $Y_B$ one can calculate the value of $\mathcal{S}_{\mathrm{weak}}$ and $\mathcal{S}^{\mathrm{ave}} = \langle \mathcal{S}_{\mathrm{weak}} \rangle$ [see Eqs. (B1)–(B5) in [2]], which is consistent with Tsirelson bound. Despite this, one cannot claim that the Bell parameter $S$, which was investigated in Bell tests, was measured on a single pair of photons. Besides, as Bohr taught us, we do not see individual photons before we register a click.

The fact that filtering-type successive measurements of incompatible observables can be performed on identically prepared physical systems is well understood and quantum predictions cannot be reproduced using Bell-local realistic coupling [23,24]. The experiment in [2], in which the successive weak measurements are performed, follows a different scenario that is tested in the Bell CHSH game. While Alice and Bob can change the ordering of the joint measurements that are performed, they can never intervene to choose to perform only one measurement setting independently of the other because the output of the first weak measurement is the input of the second weak measurement.

The experiment [2] is original and interesting. It may lead to new applications in quantum information and computing, but it neither challenges our understanding of Bell tests nor allows having doubts about Bohr complementarity and contextuality.

*Data availability.* No data were created or analyzed in this study.